\documentclass[final,5p,times,twocolumn]{elsarticle}

\usepackage{amssymb}
\usepackage{graphicx}
\everymath={\displaystyle}
\newcommand{\be}[1]{\begin{equation} \label{eq:#1}}
\newcommand{\ee}{\end{equation}}
\newcommand{\refe}[1]{(\ref{eq:#1})}

\journal{Astroparticle Physics}

\begin{document}

\begin{frontmatter}

\title{Power Spectrum Analyses of Nuclear Decay Rates}
%\title{Nuclear Decays and Solar Activity: Implications of Power Spectrum Analyses}

\author[411FLTS]{D. Javorsek II$^{\dag}$}
\address[411FLTS]{411th Flight Test Squadron, 412th Test Wing, Edwards AFB, CA 93524 USA}
\author[Stanford]{P.A. Sturrock}
\address[Stanford]{Center for Space Science Astrophysics, Stanford University, Stanford, CA 94305 USA}
\author[StJohns]{R.N. Lasenby}
\address[StJohns]{St. John's College, Cambridge, CB2 1TP, UK}
\author[Cavendish]{A.N. Lasenby}
\address[Cavendish]{Astrophysics Group, Cavendish Laboratory, Cambridge, CB3 0HE, UK}
\author[PUPhysics]{J.B. Buncher}
\address[PUPhysics]{Department of Physics, Purdue University, West Lafayette, IN 47907 USA}
\author[PUPhysics]{E. Fischbach}
\author[PUPhysics]{J.T. Gruenwald}
\author[PUPhysics,AFROTC]{A.W. Hoft}
\address[AFROTC]{Detachment 220, Air Force Reserve Officer Training Corps, West Lafayette, IN 47907 USA}
\author[PUPhysics,AFROTC,47OSS]{T.J. Horan}
\author[PUNucEng,PUPhysics]{J.H. Jenkins}
\address[PUNucEng]{School of Nuclear Engineering, Purdue University, West Lafayette, IN 47907 USA}
\author[PUPhysics,AFROTC]{J.L. Kerford}
\author[USAFA]{R.H. Lee}
\address[USAFA]{Department of Physics, United States Air Force Academy, CO 80920 USA}
\author[PUPhysics]{A. Longman}
\author[PUPhysics]{J.J. Mattes}
\author[USAFA]{B.L. Morreale}
\author[USAFA]{D.B. Morris}
\author[USAFA,NAS]{R.N. Mudry}
\address[NAS]{Training Squadron 3, Training Air Wing 5, NAS Whiting Field, Milton, FL 32570 USA}
\author[PUPhysics]{J.R. Newport}
\author[USAFA]{D. O'Keefe}
\author[PUPhysics,AFROTC,392TS]{M.A. Petrelli}
\address[392TS]{392nd Training Squadron, 30th Space Wing, Vandenberg AFB, CA 93437 USA}
\author[AFROTC]{M.A. Silver}
\author[PUPhysics,AFROTC]{C.A. Stewart}
\author[47OSS,USAFA]{B. Terry}
\address[47OSS]{47th Operations Support Squadron, 47th Flying Training Wing, Laughlin AFB, TX 78843 USA}

%\date{\today}

%----------
% Abstract
%----------
\begin{abstract}
We provide the results from a spectral analysis of nuclear decay data displaying annually varying periodic fluctuations. 
The analyzed data were obtained from three distinct data sets: $^{32}$Si and $^{36}$Cl decays reported by an experiment 
performed at the Brookhaven National Laboratory (BNL), $^{56}$Mn decay reported by the Children's Nutrition Research 
Center (CNRC), but also performed at BNL, and $^{226}$Ra decay reported by an experiment performed at the 
Physikalisch-Technische-Bundesanstalt (PTB) in Germany.
All three data sets exhibit the same primary frequency mode consisting of an annual period.
Additional spectral comparisons of the data to local ambient temperature, atmospheric pressure, relative humidity, 
Earth-Sun distance, and their reciprocals were performed.   
No common phases were found between the factors investigated and those exhibited by the nuclear decay data. 
This suggests that either a combination of factors was responsible, or that, if it was a single factor, its effects 
on the decay rate experiments are not a direct synchronous modulation. 
We conclude that the annual periodicity in these data sets is a real effect, but that further study involving additional
carefully controlled experiments will be needed to establish its origin.
%We conclude that the annual periodicity in these data sets is a real signal, 
%of as yet uncertain origin, but one which may be connected to solar activity.
\end{abstract}


\end{frontmatter}

%\linenumbers

%--------------
% Introduction
%--------------
\section{Introduction}

It is generally believed that nuclear decay rates are constant under virtually all conditions
\cite{RCE1930} and follow the familiar exponential decay law,

\be{Decay}
  \dot{N}(t)\equiv\frac{dN}{dt}=-\lambda N_0e^{-\lambda t},
\ee

\noindent
where $N(t)$ is the number of atoms remaining at time $t$, $N_0=N(0)$ is the initial
number of atoms, and $\lambda=\ln(2)/T_{1/2}$ is a constant determined by the nuclear half-life $T_{1/2}$.
However, recent work has uncovered evidence of an unexpected periodicity in several nuclear decays
\cite{AHN1986,Ellis1990,SSS1998,Shnol1998,Baurov2007}.
An examination of the periodic behavior exhibited by the data indicated that the primary frequency observed in 
each data set corresponded to a period of close to one year~\cite{Jenkins2009a,Fischbach2009}.
This observation naturally raises the possibility that either the detectors or the decays in question are 
being affected in some fashion
by an external influence such as seasonal temperature variations or solar radiation.

The reported periodic variations in decay rates motivated other experiments and analyses, which both supported
and questioned the original results~\cite{Jenkins2009b,Norman2009,Cooper2009,Jenkins2010}.
Data obtained during the solar flare of 2006 December 13 exhibited a significant dip in the counting rate of $^{54}$Mn
nearly coincident in time with a solar flare, thus supporting the suggestion of a connection between nuclear decay 
rates and solar radiation~\cite{Jenkins2009b}.
On the other hand, in an analysis of data from several decay measurements, Norman et al.~\cite{Norman2009} reported no 
evidence for a deviation from Eq. \refe{Decay}.
Neither did Cooper~\cite{Cooper2009}, in an analysis of data 
obtained from the radioisotope thermoelectric generators (RTGs) employed in the Cassini mission.
More recently a detailed analysis of the response of the detector systems employed in Refs.~\cite{AHN1986} and 
\cite{SSS1998} concludes that the annual fluctuations in decay rates reported in these references are not likely
attributable to environmental influences, such as seasonal temperature and pressure variations thereby suggesting
the possibility the decays, and not the detectors, are affected by an external influence~\cite{Jenkins2010}.

The objective of the present paper is to present a more detailed spectral analysis of the decay variations reported
earlier. 
Here we study both the decay rate frequency content and, for the first time, the 
phases of the observed signals.
In addition to verifying the annual periodicity by three different analyses, we show that these 
phases provide an important tool in discriminating among possible explanations of the
observed annual periodicities.
As we discuss below, the new phase information has two important implications: phase can have a significant impact
on the Pearson correlation coefficient, $r$, and chi-squared, $\chi_{\nu}^2$, values used by Norman et al. 
\cite{Norman2009}, and properly incorporating phases can influence the conclusions to be drawn from their data.
%Secondly our phase results indicate that the observed periodic signals are more compatible with solar rather than 
%environmental influences such as temperature or pressure.

\begin{figure*}[t]
  %\resizebox{\columnwidth}{!}{\includegraphics{CNRCPSD}} 
  \includegraphics[width=5.25in]{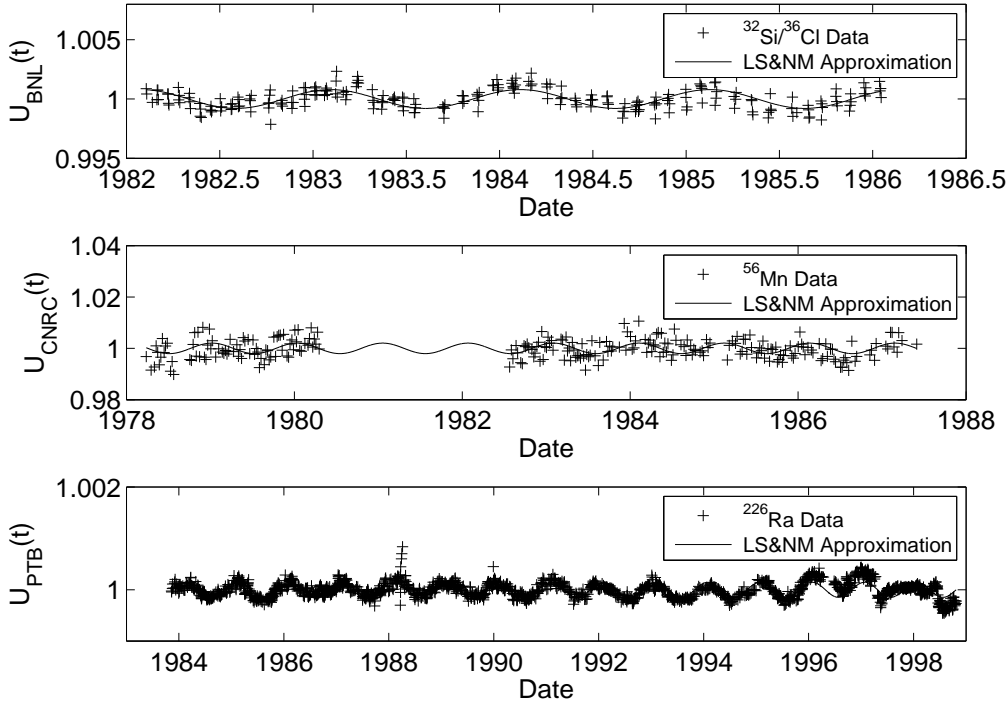}
   \caption{Reproduction of the normalized, undecayed counting rates reported by the Brookhaven National Laboratory 
\cite{AHN1986}, Children's Nutrition Research Center~\cite{Ellis1990}, and the Physikalisch-Technische-Bundesanstalt 
\cite{SSS1998}. 
These graphs include the curve fits obtained using the Lomb-Scargle (LS) and Nelder-Mead (NM) methods, as discussed in the text.}
  \label{fig:Predictions}
\end{figure*}

The first experiment analyzed was the determination of the half-life of $^{32}$Si, a $\beta^{-}$-decay, 
performed at Brookhaven National Laboratory (BNL) in Upton, New York, between 1982 and 1986~\cite{AHN1986}.  
In the BNL experiment, it was necessary for the counting rate to be directly measured over several years due to 
the long half-life of $^{32}$Si, which was determined by the BNL group to be $T_{1/2}=172\pm4$ years.  
The $^{32}$Si sample had a copper foil
which strongly attenuated the relatively low-energy $\beta$-emissions from the parent decay. 
For this reason, the subsequent decay of the higher-energy daughter species, $^{32}$P (a $\beta^{-}$-decay with a
half-life of 14.3 days), was measured and (assuming secular equilibrium) was used to infer the decay rate of 
the $^{32}$Si. 
Although insignificant for the power spectrum analysis, this decay sequence may play a role in 
the overall phase of the observed oscillations if the two nuclei are affected differently. 
As is typical for experiments of this nature, the nuclear decay of interest (in this case $^{32}$Si) was 
continuously monitored in parallel with a much longer half-life species ($^{36}$Cl) whose decay rate was assumed 
constant because of its long half-life, $T_{1/2}=301,000$ years. 
Both nuclides were measured using a gas proportional detector, which alternated between the two nuclides by use 
of a precision sample changer~\cite{AHN1986,HKW1973}. 
The BNL team noted that the ratio of the two counting rates significantly reduced apparatus-dependent systematic 
effects such as drifts in the measuring electronics, or environmental changes such as temperature, atmospheric 
pressure and relative humidity.  
Periodic annual deviations were observed in the $^{32}$Si, $^{36}$Cl counting rates, and interestingly in the ratio 
$^{32}$Si/$^{36}$Cl.
The analysis of possible laboratory temperature and humidity effects led them to conclude the deviations were of 
unknown origin.
We were able to obtain from the BNL group the original raw data and laboratory notes for the entire 
experiment~\cite{Alburger07}. 
Although the experiment ran continuously, measurements were made at semi-regular intervals that ranged from 10 
hours to 53 days with a median interval of 1 day.
This non-uniform sample rate played an important role in the spectral analysis of the data. 
See Figure \ref{fig:Predictions} for the normalized, undecayed $^{32}$Si/$^{36}$Cl counting rate.
A detailed discussion of the BNL experiment is given in Ref.~\cite{AHN1986}.

The second experiment analyzed the calibration of a $^{238}$PuBe alloy neutron total body irradiator~\cite{Ellis1990}. 
Since these data originate from an experiment utilizing the decay of $^{238}$Pu, they are directly relevant to the Cooper
analysis of $^{238}$Pu powered RTGs. 
Although the experiment was conducted (from 1977 to 1987) at BNL, it was overseen by Ellis~\cite{Ellis1990} affiliated with
the Children's Nutrition Research Center (CNRC) in Houston, Texas. 
In order to quantify the neutron flux of the $^{238}$PuBe sources used for the neutron activations, the experiment 
utilized the activation of $^{55}$Mn to $^{56}$Mn, and compared the $^{56}$Mn decays (by observing the 846.8 keV $\gamma$-ray
emitted by $^{56}$Fe) to a $^{137}$Cs standard. 
Here, the unexplained fluctuations were observed in the $^{56}$Mn decay data (a $\beta^{-}$-decay) and exhibited a 
seasonal difference of approximately 0.5\% between the winter and summer months. 
In addition to the consideration of changes in counting efficiencies, and the contribution from the low energy component 
of the neutron spectrum, all other possible causes of the seasonal variation focused on effects that would impact 
neutron emission rates.
Ultimately these were excluded since none could produce the observed variation, nor explain why no oscillations
were observed in $^{137}$Cs.
Since no known explanation could be provided for the periodic signal, a discussion of the medical impacts of subjects 
entering into treatment at different times of the year was made.
As was the case for BNL, the experiment ran continuously, but calibration measurements were made at semi-regular 
intervals that ranged from 5 to 829 days, with a median interval of 7 days, and a sizeable gap from 1980 to 1982.
See Figure \ref{fig:Predictions} for the normalized, undecayed $^{56}$Mn counting rate.
Details of the CNRC experiment are given in Ref.~\cite{Ellis1990}.

The third experiment dealt with the long-term stability of nuclear decay detectors, along with the determination of 
the half-life of europium, and was performed at the Physikalisch-Technische-Bundesanstalt (PTB) in Brunswick, Germany, 
between 1983 and 1998~\cite{SSS1998}. 
Although the raw $^{152}$Eu, $^{154}$Eu, and $^{155}$Eu data were not available, we were able to obtain data for the 
$^{226}$Ra used as the long-lived comparison standard ($T_{1/2}=1600$ yr), and these data exhibited annual periodic 
fluctuations similar to those seen in the BNL and CNRC data. 
For the ionization chamber investigations, the experimental apparatus alternated among the europium samples, 
$^{226}$Ra, and no source, in turn.
This methodology of monitoring the background during each cycle allowed background corrections to be made to the 
data directly.
Although $^{226}$Ra decays via an $\alpha$-emission, the decay chain is much more complicated than in the previously 
discussed experiments, and includes some $\beta$-emissions~\cite{CBD2007}. 
In addition, the 4$\pi\gamma$ high-pressure ionization detector used in the experimental setup reported the overall 
current produced by all the emissions in the decay chain. 
As in the BNL experiment above, the oscillation was observed in both the nuclear decays of interest (the europium samples)
and the long-lived comparison standard ($^{226}$Ra) as well as on both the ionization chamber and semiconductor detectors.
However, while the oscillation persisted in the BNL ratio, forming the ratio $^{152}$Eu/$^{226}$Ra virtually eliminated
the oscillation. 
The cause of the oscillation was attributed to a discharge effect on detector components caused by background radioactivity
due to radon and daughter products which have known seasonal fluctuations.  
During the 15 year duration of the experiment, measurements were made at intervals that ranged from 1 to 75 days with
a median interval of 3 days. 
See Figure \ref{fig:Predictions} for the normalized, undecayed $^{226}$Ra counting rate, and
Ref.~\cite{SSS1998} for the details of the PTB experiment.

%-----------------------------
% Spectral Analysis Methods
%-----------------------------
\section{Spectral Analysis Methods}

Spectral analysis techniques contain several powerful tools for the study of time series, one of which is power 
spectrum analysis 
\cite{OSB1999,Chatfield1989,HB1986}. 
In power spectrum analysis, the power, $S$, at any specified frequency, $\nu$ (measured in cycles per year), is computed 
from a time series formed from a sequence of measurements.  

Since all three experiments had non-uniform sampling rates, a simple discrete Fourier transform (DFT) could not be 
utilized. 
To deal with any problems arising from possible time-varying nuclear decay rates, three independent spectral analyses were 
carried out (which were led by DJ, PS and RL+AL), using different mathematical techniques. 
The three different techniques employed were the Lomb-Scargle method~\cite{Lomb1976,Scargle1982}, a likelihood 
procedure~\cite{Sturrock2003}, and Bayesian Spectrum Analysis~\cite{Bretthorst1988}. 
In each of these methods, the power spectrum analysis results in a Periodogram or Power Spectral Density (PSD) which 
gives the power as a function of the frequency for the data. 

One standard technique for addressing whether the data have a deterministic component is to estimate the 
``false-alarm" probability~\cite{Scargle1982,Press1992}. 
Using this method, the probability of obtaining a power $S$ or larger by chance is given by

\be{prob}
  P({\rm false \, \, alarm})=1-(1-e^{-S})^M
\ee

\noindent
where $M$ is the estimated number of independent frequencies within a desired bandwidth. 
For the false alarm probability estimate, the number of independent frequencies is taken to be the number of 
peaks observed in the bandwidth between zero and the median Nyquist frequency for each data set. 
This provides a method for identifying the statistically significant peaks in a given power spectrum. 

In addition to determining the power as a function of frequency $\nu$, the phase can be determined for a given 
frequency.  
For $\nu=1.0$ year$^{-1}$, we adopt January 1st as the zero phase reference date.
It is also important to note that the reported phase is an average of the entire data set and that the instantaneous 
phase may be variable.

When searching for common features in two or more power spectra, it is convenient to use the joint power 
statistic (JPS) which is a measure of correlation given by a function of the product of power spectra~\cite{Sturrock2005}.
The JPS has the property that, if each power spectrum has an exponential distribution (such that the probability of 
getting the power $S$ or more is $e^{-S}$), the JPS also has an exponential distribution.
For the four power spectra involved in the experiments, the JPS statistic $J$ is given, to good approximation, by
\cite{Sturrock2005}

\be{J4}
  J=\frac{3.881X^2}{1.269+X}
\ee

\noindent
where $X$ is the geometrical mean of the four powers, %$S_{Si}$, $S_{Cl}$, $S_{Mn}$, and $S_{Ra}$,

\be{X}
  X=(S_{\rm Si}S_{\rm Cl}S_{\rm Mn}S_{\rm Ra})^{1/4}.
\ee

%-----------------------------
% Power Spectrum Analysis
%-----------------------------
\section{Power Spectrum Analysis}

Before applying spectral analysis techniques it is convenient to work with the normalized, ``undecayed" data, 
$U(t)$, 

\be{NormUndecayed}
  U(t)=\frac{\dot{N}(t)}{\dot{N}(0)}e^{+\lambda t}  
\ee

\noindent
which removes the known decay trend from each data set. 
In the absence of any time-dependence other than that expected from Eq. \refe{Decay}, $U(t)$ should be constant.

\begin{figure*}[t]
  %\resizebox{\columnwidth}{!}{\includegraphics{CNRCPSD}} 
  \includegraphics[width=5.25in]{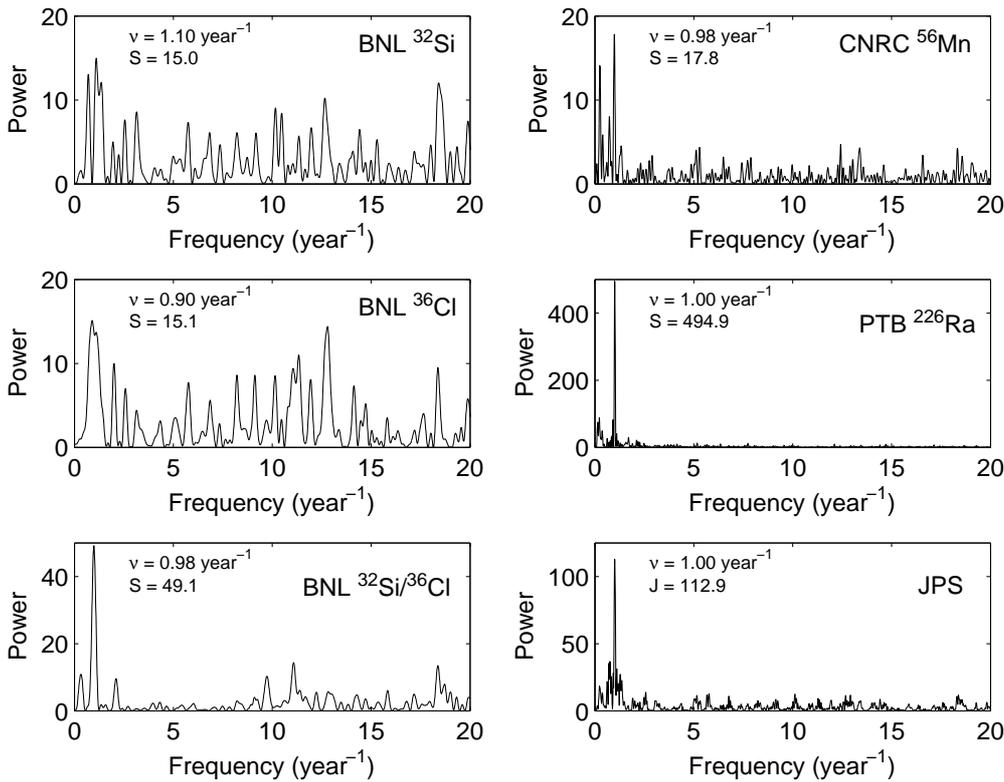}
   \caption{Power spectral densities (PSD) of the nuclear decays from the Brookhaven National Laboratory 
\cite{AHN1986}, Children's Nutrition Research Center~\cite{Ellis1990}, the Physikalisch-Technische-Bundesanstalt 
\cite{SSS1998}, and the joint power statistic (JPS) of all the nuclei. 
The frequency and power of the closest peak to an annual period are provided for each nuclide.
%The scales are set to display the annual period and low frequency spectrum behavior. 
These PSD graphs were obtained using the Lomb-Scargle method, as discussed in the text.}
  \label{fig:PSDbig}
\end{figure*}

All three mathematical techniques were applied to the BNL and PTB data, and resulted in the same general conclusion 
that the annual periodicity is a real signal for which the cause is yet to be determined. 
For reference, the PSD's from the Lomb-Scargle (LS) spectral analysis of the three experiments are provided in 
Fig.~\ref{fig:PSDbig}.  
Figure \ref{fig:PSDbig} exhibits the peak widths and amplitudes of the different nuclides and focuses on
the annual period. 
%$In addition to the definite peak near $\nu=1.0$ year$^{-1}$ which confirms what is readily apparent from 
%$Fig.~\ref{fig:Predictions}, we draw attention to the region at $\nu=10-15$ year$^{-1}$ which possesses non-aliased
%$frequency content to varying degrees for each nuclide. 
The analysis is complicated by aliasing that arises from the interplay of the very strong periodicity at $\nu=1.0$
year$^{-1}$ with the sampling time series.
Several independent methods were used to identify aliasing peaks in the spectra, and included forming the power 
spectrum of the timing, observing the effects of subtracting the annual period, and application of reference 
function comparisons.
We found that several statistically significant peaks at higher frequencies are due to the sample rate and are not 
actually present in the data. 
The frequency, power, and false alarm probability associated with the two most prominent non-aliased peaks for each 
nuclide are given in Table \ref{table:DataPSDsummary}.

%-----------------------------
% Phase Analysis
%-----------------------------
\section{Phase Analysis}

We can gain significant insight into possible causal factors when their expected frequency content and 
phase information are compared to those of the data sets. 
For example, we would expect detector systematic and environmental effects to produce the same phase for each sample in 
experiments in which the detector alternated between different samples during each run~\cite{Jenkins2010}.
From the power spectrum alone, any potential causal factor with an annual period would produce a PSD with a prominent 
peak centered at the same frequency as that observed in the experiments. 
For possible factors with direct synchronous modulation, the frequency and phase of the nuclear decay 
data would mirror that of the causal factor thereby providing possible rejection criteria. 
For reference, the phases of the four nuclides from the experiments are provided in Table \ref{table:Phases}.  

Four independent possible factors with known annual periodicities and their reciprocals were investigated. 
These included ambient air temperature at the experiment location ($T$), atmospheric pressure ($P$), relative 
humidity ($H$), and the Earth-Sun distance ($R$).
National Oceanic and Atmospheric Administration (NOAA) and United States Air Force (USAF) archived observations
from John F. Kennedy International Airport in Queens, New York (near Upton), were used for the BNL 
and CNRC experiments, while observations for Brunswick (Braunschweig) Germany, were used for the PTB comparison.

Although many possible factors possess an annual period, their phases may be significantly different.
For example, the mid-latitude temperature history of sites in both the northern and southern hemispheres would 
share the frequency content in a PSD but would be $\sim180$ degrees out of phase. 
It is important to note that the ambient atmospheric conditions discussed here are those of the conditions
outside the laboratory and are not as stringent as limits that could be obtained if the actual laboratory
environment had been monitored and reported.
For the purposes of this paper we make the assumption that these represent a conservative estimation, since
the laboratory conditions likely did not experience as dramatic fluctuations as those of the outside atmosphere.
Additionally, we assume that trends within the laboratory generally followed those of the outside
ambient conditions.
For example, we would expect that as the absolute humidity of the ambient outside air increased so too would the
conditions within the lab, albeit to a lesser degree.
Table \ref{table:FactSum} provides a summary of the power and phase for each factor.  

%It follows from Table \ref{table:FactSum} that, based on our phase analysis alone, individual environmental factors such
%as $T$, $P$, $H$, and $R$ are not likely explanations of the observed fluctuations in count rates.
%This conclusion is reinforced by a detailed discussion of the influence of environmental factors given in 
%Ref.~\cite{Jenkins2010}, which includes consideration of seasonal fluctuations in radon concentrations, which have
%been put forward as a possible explanation of the PTB data~\cite{SSS1998}.

It is natural to search for possible correlations between external influences and
the nuclear decay data using traditional parametric tests. 
We note that the relative phase between two periodic functions has a significant 
impact on goodness of fit statistics, including the Pearson correlation coefficient, $r$, and the chi-squared statistic,
$\chi^2_{\nu}$. 
For example, a periodic signal in the nuclear data with a phase different from $1/R^2$ could be masked in an analysis
comparing these data if phase information is not included.
This is evident by noting that for two periodic signals with a common frequency $\nu$, but with phases $\phi_1$ 
and $\phi_2$, $r=\cos(\phi_1-\phi_2)$ when an integral number of cycles is compared.
For a non-integer number of cycles, the expression is more cumbersome but still depends on the relative
phase and ranges from -1 to 1.
For phases on the order of those observed in the data the application of $r$ or $\chi^2_{\nu}$, as is done by
Norman et al.~\cite{Norman2009}, would result in a poor fit statistic when compared to an otherwise good periodic fit.
We show this in detail in Ref.~\cite{OKeefe2010}, where we re-examine the $^{22}$Na/$^{44}$Ti raw data of 
Ref.~\cite{Norman2009} which have been generously provided to us.
Using the power spectrum analysis methods provided above, the $^{22}$Na/$^{44}$Ti data yield a frequency of 
$1.09\pm0.33$ year$^{-1}$ with more than a 97\% confidence the peak is statistically significant
and a corresponding phase of $-25$ degrees.
When these data are analyzed incorporating the 30 day phase shift between 1/$R^2$ and the peak counting rates noted in
Refs.~\cite{AHN1986,SSS1998,Jenkins2009a,Fischbach2009}, the resulting fit has
a better $\chi^2_{\nu}$ than does the null hypothesis.

%Radon concentration fluctuations inside buildings, which increase during colder temperatures and lower pressures 
%\cite{Nazaroff1992,RS1997}, may be inferred from National Oceanic and Atmospheric Administration (NOAA) and 
%United States Air Force (USAF) archived observations of $T$, $P$ and $H$. 
%Additionally, these environmental factors may lead to density fluctuations of the gases in certain ionization 
%chambers. 
%Observations from John F. Kennedy International Airport in Queens, New York (near Upton), were used for the BNL 
%and CNRC experiments, while observations for Brunswick (Braunschweig) Germany, were used for the PTB comparison.
%It should also be noted that recent experiments constrain the possible magnitude of such effects to much less than 
%those required to cause the fluctuations realized by the BNL, CNRC, and PTB data~\cite{Abbady2004}. 
%Significantly, solar neutrino flux variations due to $1/R^2$ have been reported in the literature~\cite{Yoo2003,Hosaka2006}, 
%and additional variations have been discussed in Refs.~\cite{Sturrock1997,Sturrock2008,Sturrock2009}, 
%The fact that periodically varying signals are also observed in neutrino detectors such as Super Kamiokande, GALLEX, and
%Homestake~\cite{Sturrock2009} lends support to one of the implications of the present work, namely, that the periodic
%which supports our conjecture that the periodic
%signals observed in nuclear decay rates may also be due to periodic variations in the flux of solar nfeeutrinos.

%-----------------------------
% Discussion
%-----------------------------
\section{Discussion}

We observe from Fig.~\ref{fig:PSDbig} and Table \ref{table:DataPSDsummary} that all nuclides are dominated (to varying 
degrees) by a strong annual fluctuation. 
This was demonstrated using the Lomb-Scargle (LS) method, a likelihood procedure (LIK), and Bayesian Spectrum Analysis 
(BSA), the results of which were in good agreement.
For example, the $^{226}$Ra LS period was $363.8\pm10.1$ days, the LIK period was $365.3\pm3.3$ days, and the BSA period 
was $363.75\pm0.27$ days. 
This is also confirmed by the strong power displayed in the joint power statistic (JPS) formed from all four nuclei. 
%Additionally, %while some of the spectral content at low frequencies arises because the data are of finite duration, 
%possible low frequency content appears to exist in the range 0.2-0.4 year$^{-1}$ in $^{32}$Si/$^{36}$Cl, $^{56}$Mn, 
%$^{226}$Ra, and JPS, but not in the $^{32}$Si or $^{36}$Cl data separately. 
%However, a simple multiple-frequency model, consisting of the annual period and the lower frequency observed in the PSD, 
%is disfavored based on the results from the Bayesian spectrum analysis methods. 
%Rather there is evidence of a possible secular polynomial variation. 
%The presence of a unique polynomial baseline, coupled with 
%The fact that this low frequency is present in 
%$^{32}$Si/$^{36}$Cl but not the $^{32}$Si and $^{36}$Cl alone may provide insight into the mechanism underlying
%the observed annual fluctuation.

As noted in the Introduction, the phases of the annually varying periodic signals may help to discriminate among 
alternative explanations for the observed periodicities.
We have determined the phases of the BNL, CNRC, and PTB data sets by three different methods whose results are in close
agreement. 
Again using the $^{226}$Ra data as an example, the phases of the annual period were $-32.0\pm1.6$ deg, $-30.6\pm3.3$ deg, 
and $-30.9\pm1.1$ deg.
Representative values using the Nelder-Mead (NM) method for this phase are given in Table \ref{table:Phases}, and in Table 
\ref{table:FactSum} the phases of possible influences responsible for the periodic signals are presented.

The phases of the $^{32}$Si, $^{36}$Cl, $^{56}$Mn, and $^{226}$Ra data are different from one another, as well as also 
different from the possible factors $T$, $P$, $H$ and $R$. %indicating a possible solar influence.
To illustrate the implications of Tables \ref{table:Phases} and \ref{table:FactSum}, we note that the phases of 
$^{32}$Si and $^{36}$Cl are different, notwithstanding the fact that both nuclides were observed
in the same detector in alternating 30 minute time blocks~\cite{AHN1986}.
For example, if temperature variations at the detector were responsible for the observed periodic signals, we would expect
that the $^{32}$Si and $^{36}$Cl phases would have been the same and agree with phase of the temperature, contrary to what is observed. 
By contrast, the difference in the observed phases between $^{32}$Si and $^{36}$Cl is understandable in a picture in which the nuclides
themselves are being affected differently. %, such as by solar radiation.
We note that part of the phase difference between $^{32}$Si and $^{36}$Cl may be attributable to the fact that what is 
directly observed in $^{32}$Si are the $\beta$-emissions from the decay of daughter nuclei $^{32}$P produced by $^{32}$Si.
Since the half-life of $^{32}$P is 14.28 days, 
account must be taken in analyzing decay data of a possible phase lag between a change in the $^{32}$Si decay rate, and
its manifestation via $^{32}$P decay.
%it reaches secular equilibrium with $^{32}$Si, provided that the decay rate
%of $^{32}$Si is itself constant in time.
%It is then straightforward to show that in the presence of a time-varying influence on $^{32}$Si and/or $^{32}$P, a phase shift
%will arise between the $^{32}$Si-$^{32}$P $\beta$-emissions, and those produced by $^{36}$Cl.
%There are in addition, other features in the phase data which point to a solar influence on the decays, rather than to an
%effect on the apparatus, and these will be discussed elsewhere.
Specifically, if the dominant effect of an external influence on the decay sequence: 
${}^{32}{\rm Si} \rightarrow {}^{32}{\rm P} + {\rm e}^- + \overline{\nu}_e$, is on the parent $^{32}$Si nucleus, then any 
modification in the $^{32}$Si
decay rate will be reflected in the $^{32}$P decay rate (which is what is observed) $\sim$ 20.6 days ($={\rm T}_{1/2}$) later.
This illustrates how a mechanism which focuses on the decay process itself (rather than on the behavior of the detection 
system) could account for the different phases that are observed in the data. 
\section{Conclusions}

Following a spectral analysis of data from three separate nuclear decay experiments ($^{32}$Si and $^{36}$Cl from BNL 
\cite{AHN1986}, $^{56}$Mn from CNRC~\cite{Ellis1990}, and $^{226}$Ra from PTB~\cite{SSS1998}), we conclude from three
separate analyses that all exhibit a primary mode with an annual period. 
In addition, we determined the phase of the various signals, and found a significant phase shift in each data set 
relative to that from four possible external factors.
We have shown that ambient outdoor air temperature, atmospheric pressure, relative humidity, Earth-Sun distance,
and their reciprocals could each be excluded as the sole explanation of the observed annual fluctuations, assuming
their effects on the decay rate experiments are in direct synchronous modulation.
Although we are unable to use this analysis to prove a given hypothesis, in light of known environmental detector 
effects~\cite{Jenkins2010}, and the observation that phase shifts can exist in some experiments when nuclei are affected 
differently, we conclude that these results are consistent
with the hypothesis that nuclear decay rates may be influenced by some form of solar radiation. 
%From our spectral analysis, a mechanism which produces annual variations affecting nuclei differently is favored and
%consistent with the hypothesis that nuclear decay rates may be influenced by some form of solar radiation. 

The inference from our analysis that different nuclei may be affected differently by an external source could help to explain
recent papers by Norman et al.~\cite{Norman2009} and Cooper~\cite{Cooper2009} who have set limits on possible variations
in the decay rates of several nuclides.
Since the phases of the various nuclides summarized in Table \ref{table:Phases} differ sufficiently from 
that of $1/R^2$, failure to take them into account masks the statistically significant annual variation present in the 
data of Norman et al.~\cite{Norman2009,OKeefe2010}.
%Moreover, none of the nuclides studied in Ref.~\cite{Norman2009} coincide with those studied here,
%and hence there is no direct conflict between our results and Ref.~\cite{Norman2009}. 
More generally, it is reasonable to suppose that the same complex details of nuclear structure (e.g. nuclear wavefunctions,
angular momentum selection rules, etc.), which are responsible for the fact that half-lives vary from fractions of a second
to tens of billions of years, could also affect the response of different nuclei to some external influence.
With respect to the analysis of Cooper~\cite{Cooper2009} we have noted that since
the data of Ellis~\cite{Ellis1990} described above also derive from a $^{238}$Pu source, it is useful to compare these two 
data sets which appear to be in disagreement with each other.
We observe that in an RTG the decay of $^{238}$Pu produces $^{234}$U which has a half-life of $2.46\times10^5$ year, and hence 
an RTG is only sensitive to $\alpha$-decays.
In contrast, the $^{238}$PuBe source used by Ellis was monitored using the $\beta$-decay of $^{56}$Mn.
Absent a clear understanding of the cause of the oscillations, we simply observe that the data of Ellis~\cite{Ellis1990} and 
Cooper~\cite{Cooper2009} could be compatible in a picture in which an external influence predominantly affects $\beta$-decays.
We conclude from the preceding discussion that there is no obvious conflict between our results and those of Norman 
et al.~\cite{Norman2009} or Cooper~\cite{Cooper2009}, given our limited understanding of the mechanism underlying the annual
fluctuations discussed in the present paper.
Additional discussion of these papers can be found in Ref.~\cite{Jenkins2010}. 

Our analysis can be extended to apply to other possible explanations for the observed annual fluctuations.
As the sensitivity to nuclear decay rates improves, additional care must be taken to monitor room atmospheric conditions, 
radon levels, and power-source fluctuations, as well as the Earth's electric and magnetic fields. 
It is also possible that some mechanisms may be excluded by performing experiments where any or all are intentionally varied. 
For example, repeating one of these experiments at a mid-southern latitude should flip the phase of many atmospheric factors. 
We thus encourage the development of experiments to reproduce the observed oscillations in the specific nuclei analyzed 
here with an emphasis on carefully controlling the external environment.

%------------------
% Acknowledgements
%------------------
\section{Acknowledgements}

The authors are deeply indebted to D. Alburger, K. Ellis, G. Harbottle, E. Norman, and H. Schrader for providing us with 
their available data and D. Krause for helpful conversations.
Special thanks to the 14th Weather Squadron, 2nd Weather Group, USAF for their support in obtaining detailed weather data 
from Houston, New York, and Braunschweig, Germany.
We also wish to thank the US Air Force Academy Center for Innovation and the Air Force Office of Scientific Research for 
their ongoing support.

The views expressed in this paper are those of the authors and do not reflect the official policy or position of the U.S. 
Air Force, U.S. Department of Defense, or the U.S. Government. 
The material in this paper is Unclassified and approved for public release: distribution is unlimited, reference Air Force Flight 
Test Center Public Affairs reference number AFFTC-PA No: AFFTC-PA-09458.
The work of P.S. is supported by National Science Foundation grant No. AST-0607572.
The work of E.F. was supported in part by the U.S. Department of Energy under Contract No. DE-AC02-76ER071428.

\noindent
$^\dag$ Corresponding Author

% Table of Data PSD summary
\begin{table*}[b]
\caption{Spectral content summary of the BNL, CNRC, and PTB nuclear decay data. 
Only the two most prominent non-aliased peaks are included for each nuclide, and $^{32}$Si/$^{36}$Cl denotes the 
ratio of $\dot{N}(Si)/\dot{N}(Cl)$.} 
%The 3 dB beamwidth was used to determine the reported frequency errors.}
\label{table:DataPSDsummary}
\begin{tabular}{ccccc}
    \hline\hline
    Experiment & Nuclides & Frequency [year$^{-1}$] & Power & False Alarm Probability\\ 
    \hline
    BNL & $^{32}$Si/$^{36}$Cl & $0.98\pm0.12$ & 49.1 & $2.7\times10^{-18}$ \\
    BNL & $^{36}$Cl & $0.90\pm0.31$ & 15.1 & $1.9\times10^{-3}$ \\
    BNL & $^{32}$Si & $1.10\pm0.23$ & 15.0 & $1.9\times10^{-3}$ \\ 
    BNL & $^{32}$Si & $0.70\pm0.10$ & 13.0 & $1.3\times10^{-1}$ \\
    BNL & $^{36}$Cl & $11.34\pm0.28$ & 11.0 & $9.5\times10^{-2}$ \\
    BNL & $^{32}$Si/$^{36}$Cl & $0.33\pm0.12$ & 10.9 & $1.1\times10^{-1}$ \\
    \hline
    CNRC & $^{56}$Mn & $0.98\pm0.04$ & 17.8 & $1.6\times10^{-5}$ \\
    CNRC & $^{56}$Mn & $0.25\pm0.04$ & 14.1 & $6.4\times10^{-4}$ \\
    \hline
    PTB & $^{226}$Ra & $1.00\pm0.03$ & 494.9  & 0 \\
    PTB & $^{226}$Ra & $0.22\pm0.04$ & 82.1 & 0 \\
    \hline
    JPS & All & $1.00\pm0.04$ & 112.9 & 0 \\
    JPS & All & $0.76\pm0.06$ & 36.8 & $3.7\times10^{-13}$ \\
    \hline\hline
\end{tabular}
\end{table*}

% Table of Phase Summary
\begin{table*}[b]
\caption{Primary mode phase analysis summary using a 1.00 year$^{-1}$ frequency for
the three experiments. By convention, a phase shift of -41.4 deg corresponds to a maximum count rate
approximately 41 days after January 1.
The indicated phases represent the average over the entire data set.}
\label{table:Phases}
\begin{tabular}{ccc}
    \hline\hline
    Experiment & Nuclide & Phase [deg] \\
    \hline
    BNL & $^{32}$Si & $36.4\pm24.1$ \\
    BNL & $^{36}$Cl & $106.3\pm7.0$ \\
    BNL & $^{32}$Si/$^{36}$Cl & $-41.4\pm2.7$ \\
    CNRC & $^{56}$Mn & $16.2\pm1.1$ \\
    PTB & $^{226}$Ra & $-32.0\pm1.6$ \\ 
    \hline\hline
\end{tabular}
\end{table*}

% Table of factors
\begin{table*}[b]
\caption{Phase results of the spectral analysis for the primary modes of environmental factors by experiment. 
As in Table \ref{table:Phases}, a 1.00 year$^{-1}$ frequency was used and the indicated phases represent the average 
for the time interval corresponding to the associated nuclear decay data.
The phases of the environmental factor reciprocals are provided in the parentheses.}
\label{table:FactSum}
\begin{tabular}{c|cc|cc|cc}
    \hline\hline
     Causal & \multicolumn{2}{c|}{BNL} & \multicolumn{2}{c|}{CNRC} & \multicolumn{2}{c}{PTB} \\ 
     Factor & Power & Phase [deg] & Power & Phase [deg] & Power & Phase [deg] \\ 
    \hline
    $T$ ($1/T$) & 602.7 & 153.8 (-26.2) & 766.6 & 155.0 (-25.0) & 1920.0 & 160.2 (-19.8) \\ 
    $P$ ($1/P$) & 64.3 & 49.0 (-131.0) & 80.2 & 48.3 (-131.7) & 13.1 & 15.6 (-164.4) \\
    $H$ ($1/H$) & 34.3 & 129.4 (-50.6) & 38.4 & 132.9 (-47.1) & 761.8 & 16.3 (-163.7) \\
    $R$ ($1/R$) & 713.3 & 176.7 (-3.3) & 1675.2 & 178.4 (-1.6) & 2730.7 & 178.3 (-1.7) \\ 
    \hline\hline
\end{tabular}
\end{table*}

%Fig Predictions

%Fig PSD big

%Fig CumSum
%\begin{figure*}[t]
%  \begin{tabular}{cc}
%  \psfig{file=CumSumClTemp.ps,width=2.6in} & \psfig{file=CumSumClACRIM.ps,width=2.6in} \\
%  (a) & (b) \\ 
%  \end{tabular}
%  \caption{The cumulative summation comparisons of $^{36}$Cl with (a) temperature and (b) ACRIM.
%The similarity between the ACRIM and nuclear decay data suggests these results are compatible with the hypothesis that
%nuclear decay rates are influenced by solar activity.}
%  \label{fig:CumSum}
%\end{figure*}

\end{document}